\documentclass[12pt]{iopart}

\usepackage{iopams}
\usepackage{graphicx, listings}
\usepackage{bbold}

\usepackage[latin1]{inputenc}
\usepackage{color}
\usepackage{hyperref}
\usepackage{latexsym}

\expandafter\let\csname equation*\endcsname\relax
\expandafter\let\csname endequation*\endcsname\relax
\usepackage{amsmath}

\definecolor{drkgreen}{rgb}{0.0, 0.5, 0.0}

\begin{document}

\title[]{Multiworm algorithm quantum Monte Carlo}

\author{F. Lingua $^{1}$, B. Capogrosso-Sansone $^{1}$, A. Safavi-Naini $^{2}$, A. J. Jahangiri $^{3}$ and V. Penna $^{4}$}

\address{%
$^{1}$ \quad Department of Physics, Clark University, Worcester, Massachusetts 01610, USA\\
$^{2}$ \quad JILA, National Institute of Standards and Technology and Department of Physics, University of Colorado, 440 UCB, Boulder, CO 80309, USA\\
$^{3}$ \quad H. L. Dodge Department of Physics and Astronomy, The University of Oklahoma, Norman, Oklahoma 73019, USA\\
$^{4}$ \quad Department of Applied Science and Technology and u.d.r. CNISM, Politecnico di Torino, I-10129 Torino, Italy}

\ead{flingua@clarku.edu}

\begin{abstract}
We review the path-integral quantum Monte Carlo method and discuss its implementation by multiworm algorithms. We analyze in details the features of the algorithms, and focus our attention on the computation of the $N$-body density matrix to study N-body correlations. Finally, we demonstrate the validity of the algorithms on a system of dipolar bosons trapped in a stack of $N$ one-dimensional layers in the case of zero and finite inter-layer hopping.
\end{abstract}

%Uncomment for PACS numbers title message
%\pacs{00.00, 20.00, 42.10}
% Keywords required only for MST, PB, PMB, PM, JOA, JOB?
%\vspace{2pc}
%\noindent{\it Keywords}: Article preparation, IOP journals
% Uncomment for Submitted to journal title message
%\submitto{\JPA}
% Comment out if separate title page not required
\maketitle

\section{Introduction}

Many-body strongly-correlated quantum systems exhibit a great variety of interesting phenomena and, under certain conditions, may stabilize exotic quantum phases of matter \cite{revChen,revBloch,emergentManyBody}.
%(superfluidity, Mott-insulator, many-body localization, super-solidity\dots).
These systems attract a great deal of attention due to their wide range of potential applications, spanning from quantum computation and information tasks \cite{TopQcomp_Rev,TrappdIonsQcomp_Rev} to purely theoretical queries on fundamental laws of physics \cite{Ketterle_2009}-\cite{Capellaro_2017}.
From a theoretical viewpoint, the understanding of these systems is especially challenging unless one considers weakly- or strongly-interacting regimes where approximations can be made. Therefore, theorists often resort to advanced computational techniques capable to capture the many-body correlations in {\em any} regime. % featuring their rich phenomenology.
%Within the context of quantum simulation, Path Integral Quantum Monte Carlo techniques play a prominent role.
Within this context, Path Integral Quantum Monte Carlo techniques \cite{Barker_1979}-\cite{PIMC_2015} play a prominent role in the study of many-body bosonic systems.
%Within the context of quantum simulation, Path Integral Quantum Monte Carlo (PIMC) techniques play a prominent role, proving to be a reliable effective tool capable of extracting information on the many-body correlations.

In this paper, we first review the general scheme for Path Integral Quantum Monte Carlo technique and discuss its implementation with Multiworm algorithms suitable to study multi-component bosonic systems. We then demonstrate the validity of the algorithms by considering dipolar bosons trapped in a stack of $N$ one-dimensional layers in the case of zero and finite inter-layer hopping.
%We then show how to extend the Worm algorithm to its Many-Worms generalization in order to ensure ergodicity in the computation of many-body density matrices, and allow the study of $N$-body correlation functions among both distinguishable and indistinguishable particles. We finally provide a couple of example of application of the $N$-Worms algorithm considering a system composed by a dipolar-gas in a stack of $N$ one-dimensional layers.

\section*{Path-integral quantum Monte Carlo}
Within the formalism of quantum statistical mechanics, the expectation value of physical observables can be evaluated according to the expression
\begin{equation}
\langle \hat{O}\rangle = Tr(\hat{\rho} \hat{O})=\sum_\alpha \langle \alpha|\hat{\rho} \hat{O}|\alpha \rangle \label{expO}
\end{equation}
where $\hat{O}$ is the quantum-operator corresponding to the physical observable $O$, $\hat{\rho}=e^{-\beta\hat{H}}/\mathcal{Z}$ is the density operator, the set $\{ | \alpha \rangle \}$ form a basis for the physical states relevant  to $H$, and
\begin{equation}
\mathcal{Z}=Tr(e^{-\beta\hat{H}}) \label{partf}
\end{equation}
is the partition function.
Here, the parameter $\beta=1/k_BT$ is the inverse temperature and $\hat{H}$ is the Hamiltonian of the system.
Expectation value in Eq. (\ref{expO}) and partition function in Eq. (\ref{partf}) can be computed exactly only in very simple or non-interacting cases. When dealing with strongly-correlated quantum systems, quantum Monte Carlo simulations prove to be the most powerful technique to compute Eq. (\ref{expO}) and Eq. (\ref{partf}). In the absence of sign-problem, estimates of Eq. (\ref{expO}) and Eq. (\ref{partf}) can be achieved with controllable error-bars.
%Both the expectation value (\ref{expO}) and the partition function (\ref{partf}) can be estimated by means of analytic methods \cite{an_met} or numerical methods \cite{num_met}. However, when dealing with many-body quantum problems
%, such as cold-atoms in optical lattices and strongly correlated systems in general,
%due to the huge dimension of the Hilbert space, standard computational techniques may lack on reliability or efficiency.
%In these contexts Monte Carlo (MC) simulations prove to be a remarkable effective tool. They are capable of providing a viable mean of computation, thanks which, almost exact estimates of (\ref{expO}) and (\ref{partf}) can be achieved.

For bosonic systems, one of the most explored class of Monte Carlo techniques is the so-called path-integral Quantum Monte Carlo (PIMC). PIMC algorithms rely on the path-integral representation of the partition function $\mathcal{Z}$ \cite{FeynmStatM}, where the density operator can be treated as a unitary evolution operator in imaginary-time $\tau=i\cdot t$ with $\tau \in [ 0, \beta]$.

In this paper, we consider bosonic lattice systems described by Bose-Hubbard (BH)-type models. A generic BH model is described by the Hamiltonian
\begin{equation}
\hat{H}=-\sum_{i,j} J_{ij} a^\dag_i a_j + \sum_{i,j}V_{ij}n_i n_j + \sum_i U_i n_i(n_i - 1) -\sum_i \mu_i n_i\label{BH}
\end{equation}
where $a^\dag_i$ and $a_j$ are the creation and annihilation operators on lattice sites $i$ and $j$ respectively, satisfying the bosonic commutation relations $[a_i,a^\dag_j]=\delta_{ij}$, $n_i=a^\dag_i a_i$ is the occupation number operator of lattice site $i$, $U_i$ is the particle-particle on-site interaction, $J_{ij}$ the tunneling amplitude, $V_{ij}$ is the density-density interaction between sites $i$ and $j$, and
$\mu_i$ is the chemical potential at the site $i$.
When studying BH models, a convenient basis set is given by Fock states which are defined in the discrete spatial mode representation (i.e. Wigner basis). In this representation, the state of the system is described by a collection of occupation numbers referring to the number of particles located at each discrete position in space (i.e. lattice sites).

Hamiltonian in Eq. (\ref{BH}) can be conveniently split into two parts:
\begin{equation}
\hat{H}=\hat{H}_0 + \hat{H}_1,
\end{equation}
where $\hat{H}_0=\sum_{i,j}V_{ij}n_i n_j + \sum_i U_i n_i(n_i - 1) -\sum_i \mu_i n_i$ is the diagonal part in the chosen Fock representation, while $\hat{H}_1=-\sum_{i,j} J_{ij} a^\dag_i a_j$ is the off-diagonal part.

According to the path-integral representation of quantum mechanics \cite{FeynmStatM}, the expectation value of a generic observable $\langle \hat{O}\rangle$ can be computed as an unitary evolution in imaginary-time between $\tau=0$ and $\tau=\beta$. In the interaction picture, the trace $Tr(e^{-\beta\hat{H}}\hat{O})$ can be computed as the sum of all the possible evolutions, i.e. paths, from state $|\Theta_\alpha\rangle=\hat{O}|\alpha\rangle$ at $\tau=0$, to state $|\alpha\rangle$ at $\tau=\beta$ as follows:
%\begin{equation}
% Tr(e^{-\beta\hat{H}}\hat{O})=\sum_\alpha \sum_{paths}p_\alpha\langle \alpha| \hat{H}_{1}(\tau_{n})\cdots\hat{H}_{1}(\tau_1)|\Theta_\alpha\rangle,
% \label{pathint}
%\times\langle \alpha| e^{-\beta\hat{H}_0} \hat{K}_{l_{n}}(\tau_{n})\hat{K}_{l_{n-1}}(\tau_{n-1})\cdots\hat{K}_{l_{1}}(\tau_1)|\Theta_\alpha \rangle
%\end{equation}
\begin{equation}
 Tr(e^{-\beta\hat{H}}\hat{O})=
 \sum_\alpha \sum_{paths}p_\alpha\langle \alpha| \hat{H}_{1}(\tau_{n})|\alpha_n\rangle\cdots\langle\alpha_1|\hat{H}_{1}(\tau_1)|\Theta_\alpha\rangle,
\label{pathint}
\end{equation}
where the product of ``hopping amplitudes'' $\langle \alpha| \hat{H}_{1}(\tau_{n})|\alpha_n\rangle\cdots\langle\alpha_1|\hat{H}_{1}(\tau_1)|\Theta_\alpha\rangle$ define the single path of the many-body state in imaginary-time
 and $\hat{H}_{1}(\tau)$ is the off-diagonal part of the Hamiltonian in the interaction picture. %, and $p_\alpha$ is the probability of the path}.
Notice that for $\hat{O}=\mathbb{1}$, $|\Theta_\alpha\rangle = |\alpha\rangle$, and trace in Eq. (\ref{pathint}) is the partition function in Eq. (\ref{partf}).
For a complete review of the derivation of the path-integral formulation in the interaction picture and continuous imaginary-time, we refer to the Appendix \ref{app}.
%\rd{WE NEED TO DISCUSS about meaning and notation of path in (5) and Appendix. I propose the following modification of (5):
%\begin{multline}
%Tr(e^{-\beta\hat{H}}\hat{O})=\\
%\sum_\alpha \sum_{paths}p_\alpha\langle \alpha| \hat{H}_{1}(\tau_{n})|\alpha_n\rangle\cdots\langle\alpha_1|\hat{H}_{1}(\tau_1)|\Theta_\alpha\rangle,
%\nonumber
%\end{multline}
%}

\begin{figure}
\centering
\includegraphics[width=0.8\columnwidth]{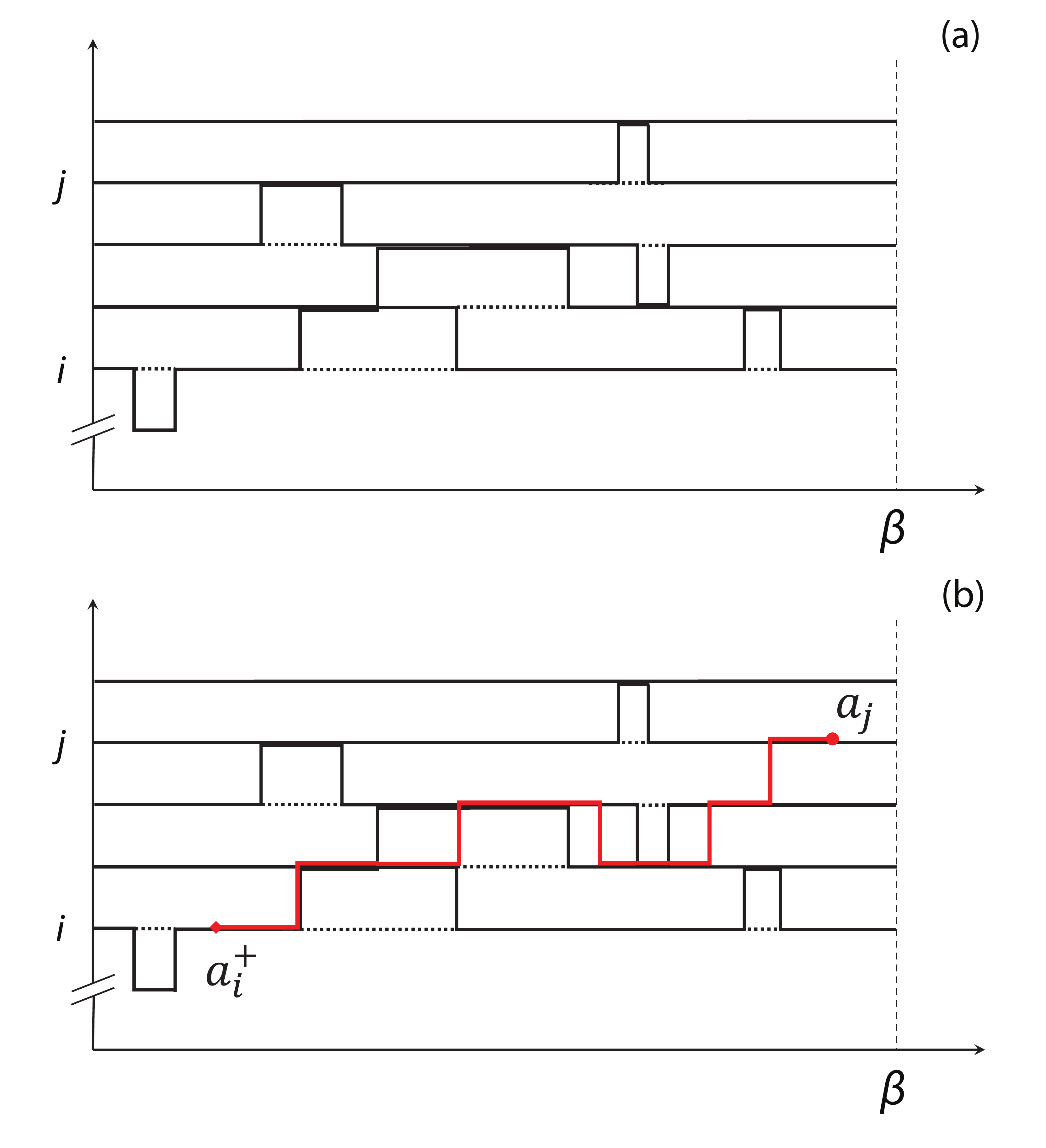}
\caption{(Color online) Typical PIMC configurations for the Bose-Hubbard model: (a) in absence of disconnected
worldlines, (b) with one (red) disconnected worldline.}
\label{fig1}
\end{figure}

The collection of the infinitely-many possible paths provide the configuration space within which the PIMC algorithm performs updates. Each configuration represents a specific path as an evolution of initial Fock state $|\alpha\rangle$ in imaginary-time.
%, and provide a description of the evolution of the Fock state of the system in imaginary-time. %When dealing with Bose-Hubbard type models, the Fock states are defined in the discrete spatial mode representation (i.e. Wagner basis). In this representation the state of the system is described by a collection of occupation numbers referring to the number of particles located at each discrete position in space (i.e. lattice sites).
%A single configuration represents a possible evolution of the particles in space and imaginary-time.
A typical example of configuration is depicted in Fig. \ref{fig1}a) where on the horizontal axis is the imaginary-time and on the vertical axis are lattice sites.
% \begin{figure}
% \centering
% \begin{minipage}{1.0\columnwidth}
% \includegraphics[width=\columnwidth]{worldline_v3up.pdf}
% \end{minipage}
% \begin{minipage}{1.0\columnwidth}
% \includegraphics[width=\columnwidth]{worldline_v3dwn.pdf}
% \end{minipage}
% \caption{(Color online) Typical PIMC configurations for the Bose-Hubbard model: (a) in absence of disconnected
% worldlines, (b) with one (red) disconnected worldline.}
% \label{fig1}
% \end{figure}
The configuration can be seen as a collection of world-lines. Each world-line closes on itself, due to the periodic boundary conditions in imaginary-time resulting from the trace operation, and represents a single particle propagating in imaginary-time and space.
%The points in imaginary time where the system changes state are called kinks.

Formally, (\ref{expO}) and (\ref{partf}) can be written as $\mathcal{Z}=\sum_\nu W_\nu$ and $\langle O\rangle=\sum_\nu O_\nu W_\nu/\mathcal{Z} $ respectively, where $W_\nu$ is the weight of each configuration, the index $\nu$ is a collection of discrete and continuous indexes (see Appendix \ref{app} for further details) labelling a specific configuration, and $O_\nu$ is the value assumed by observable $O$ in configuration $\nu$. Summing over all possible configurations is practically impossible. The solution is provided by the Metropolis method \cite{MetMed} according to which configurations are sampled with a probability proportional to their weight $W_\nu$. At each Monte Carlo step a different configuration proposed via some updating procedure is accepted or rejected with a probability proportional to $W_\nu$ according to detailed balance equation \cite{prkof_1,prkof_2}.

%Without loss of generality, the partition function (i.e. and similarly the expectation value) can be formally written as $\mathcal{Z}=\sum_\nu W_\nu$ ($\langle O\rangle=\sum_\nu O^\nu W_\nu/\mathcal{Z} $), where $W_\nu$ is the weight of each configuration and the index $\nu$ runs over all the possible configurations (see Appendix \ref{app} for further details).
%However, in continuous time the configuration space is infinitely-large, and it is obviously impossible to sum over all possible paths. On the other hand, the Monte Carlo algorithm cannot just uniformly-random sample the configuration space; with near certainty the selected configurations would be having vanishing small weights, and the algorithm would result largely inefficient. A possible solution is to sample only those configuration that contribute the most to the final partition function. This is achieved by sampling configurations with a probability proportional to their weight $W_\nu$ via the so called Metropolis Method \cite{MetMed}. At each Monte Carlo step a different configuration is generated, the move is then accepted or rejected with a probability proportional to $W_\nu$, that satisfies a proper detailed-balance equation \cite{prkof_1,prkof_2}. It is important to stress that in deriving the weight of a configuration, no approximations are introduced (see Appendix \ref{app} for details).

\subsection{Worm algorithm}

The Worm algorithm~\cite{prkof_1,prkof_2}, developed by Prokof'ev and al. , is a PIMC technique that works in an enlarged configuration space where a disconnected world-line, the worm, is allowed (red line in Fig. \ref{fig1}). Configurations containing worms are generated by a generalized Hamiltonian $\hat{H}\rightarrow \hat{H} -\gamma \hat{Q}$, where, $\gamma$ is a coefficient which can be chosen in order to optimize the efficiency of the algorithm, and the source term $\hat{Q}$ has the form:
\begin{equation}
\hat{Q}(\tau)=\sum_{i}[ a_i^{\dag}(\tau) + a_i(\tau)],
\label{source1}
\end{equation}
where $a_i(\tau)=e^{\tau\hat{H}_0}a_i e^{-\tau\hat{H}_0}$ ($a_i^{\dag}(\tau)=e^{\tau\hat{H}_0}a^\dag_i e^{-\tau\hat{H}_0}$) are the annihilation (creation) operator at site $i$ expressed in the interaction representation.
In the path integral formulation described in Appendix \ref{app}, the term $\hat{Q}(\tau)$ is added to the off-diagonal term $\hat{H}_{1}(\tau)$.
When the expansion procedure (see Appendix \ref{app}) is applied to $\mathbf{\hat{T}}e^{-\int_0^\beta\hat{H}_1(\tau) + \hat{Q}(\tau)\, d\tau}$, terms with only one annihilation (creation) operator at different positions $i$ ($j$) and times $\tau$ ($\tau^\prime$) appear in the expansion. These terms correspond to configurations with multiple disconnected worldlines, i.e. worms, where ``head'' and ``tail'' of a worm correspond to the annihilation operation $a_j(\tau)$ and the creation operator $a_i^\dag(\tau)$ respectively. In the lower panel of Fig. \ref{fig1}, we show an example of configuration containing a single worm (red line). The local action of $a_i^\dag$ (tail) and $a_j$ (head) on the configuration increases the particle number of $+1$ on site $i$ at time $\tau$, and decreases it of $-1$ on site $j$ at time $\tau^\prime$.
%Notice that, by expanding the term \bl{$\mathbf{\hat{T}}e^{-\int_0^\beta\hat{H}_1(\tau) + \hat{Q}(\tau)\, d\tau}$}, configurations with multiple worms are generated.
For the sake of simplicity, unless otherwise needed, only configurations with one worm are considered. We shall see how, in many cases, to ensure ergodicity, multiple worms need to be included.

At each Monte Carlo step, a new configuration, obtained via a certain updating procedure, is proposed. Within the Worm algorithm, all updates but one (the create-worm update, see below) happen by moving head or tail of a worm~\cite{prkof_1,prkof_2}. These configurations correspond to terms obtained by expanding $\mathbf{\hat{T}}e^{-\int_0^\beta\hat{H}_1(\tau) + \hat{Q}(\tau)\, d\tau}$ ( see Appendix \ref{app}).
FIG. \ref{fig_update} (a)-(c) shows three updating procedures and corresponding counter-updates (see caption for details). All updates are local, i.e. they change the configuration on a local region in space and imaginary-time. Reading from left to right, panel (a) shows in the top (bottom) right sketch the annihilation (creation) of a worm with its head (tail) and tail (head)
at $\tau=\tau_1$ and $\tau=\tau_2$ respectively.
%{\color{red} I am not sure this is clear, I do not know ow else to write it}
%\bl{shows the creation of a worm (a single worm appears in the configuration) with its head and tail at times $\tau=\tau_1$ and $\tau=\tau_2$ in the top-right sketch  of FIG.~\ref{fig_update} (a), or tail and head at times $\tau=\tau_1$ and $\tau=\tau_2$ in the bottom-right sketch.}
The former corresponds to erasing a piece of an existing worldline, the latter corresponds to drawing a piece of a new worldline. Panel (b) shows the \emph{shift-in-time} update where the head of the worm is shifted in imaginary-time from $\tau_1$ to $\tau_2$. Panel (c), reading from left to right, shows the  \emph{kink} update where the head of the worm is shifted in space from site $i$ to site $j$.
\begin{figure}
\centering
\includegraphics[width=\columnwidth]{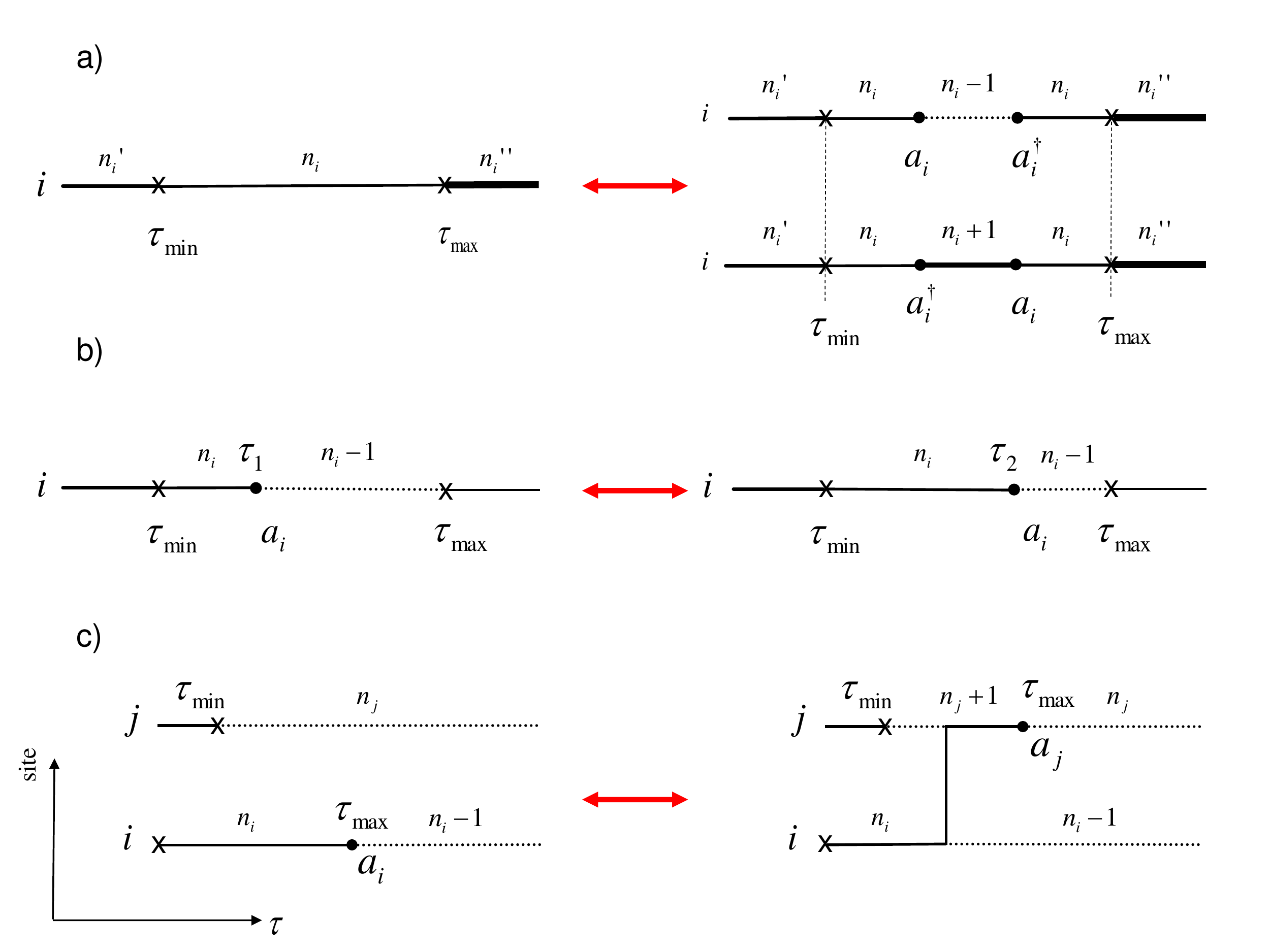}%update.eps
\caption{Three updating procedures (reading the figure from left to right) and corresponding counter-updates (reading the figure from right to left) on an imaginary-time interval $[\tau_{min}, \tau_{max}]$ of the configuration. Thickness of the line is proportional to the occupation number $n_i$ of the site $i$ in that time interval. Dashed line means $n_i=0$.
%(a) creation of a worm on lattice site $i$, with its head and tail at imaginary-time $\tau=\tau_1$ and $\tau=\tau_2$ (top-right sketch), head at $\tau=\tau_2$, and tail at $\tau=\tau_1$ (bottom-right sketch);
(a) in the top-right sketch (bottom-right sketch) creation of a worm on lattice site $i$, with its head (tail) at imaginary-time $\tau=\tau_1$ and tail (head) at imaginary-time $\tau=\tau_2$;
(b) shift in time of the head of the worm from $\tau=\tau_1$ to $\tau=\tau_2$;
(c) jump in space from site $i$ to site $j$ of the head of the worm. }
\label{fig_update}
\end{figure}
%(jump, reconnection, shift in time \dots \cite{prkof_1,prkof_2}) %Among all the possible actions that updating procedures can take, are also those that remove head and tail of the worm from the configuration.

Overall, there exist two classes of configurations: those in which the worm is present (upper panel of Fig. \ref{fig1}), and those in which it is not (lower panel of Fig. \ref{fig1}).
%The latter will be used to estimate the expectation value of a given operator $\hat{O}$ according to the relation (\ref{expO}), while the former will be used to collect statistics for the estimate of the Green function.
Since operator $\hat{Q}(\tau)$ does not belong to the original BH Hamiltonian, configurations in which the worm is present cannot be used to compute the expectation value of any physical observable $O$. On the other hand, when the worm is present, configurations can be used to collect statistics for the Green function which, in the interaction picture, is defined as:
%the trace operation, by definition, has to be computed between the same initial and final state $|\alpha\rangle$. In other words, being $O$ a physical observable the overall action of the associated operator $\hat{O}$ and its imaginary-time evolution has to be hermitian. Any action of creation and annihilation operators modifying the occupation number on different sites, at different imaginary-times, would cause a final Fock state different from the initial one, causing a non-hermitian update of the estimate of $\langle\hat{O}\rangle$.
%However, when the worm is present, the configurations will be used to collect statistics for the estimate of the Green function.
\begin{equation}
G(\vec{x}_i,\vec{x}_j,\tau,\tau^\prime)=\langle \mathbf{\hat{T}}_\tau a_j(\tau^\prime)a^\dag_i (\tau)\rangle,
\label{GF}
\end{equation}
where $\mathbf{\hat{T}}_\tau$ is the time-ordering operator. Indeed, when (\ref{GF}) is expressed using path integral formalism as discussed in Appendix~\ref{app}, terms corresponding to configurations
%with a single worm are generated.
with a single worm appear in the expansion due to the presence of $a_j$ and $a^\dag_i$ in in Eq. (\ref{GF}).
Note that the Green function coincides with the $1$-body density matrix
%which can give access to the estimate of different quantum correlators.
%\begin{equation}
$D_1(\vec{x}_i,\vec{x}_j,\tau,\tau^\prime) \equiv \,G(\vec{x}_i,\vec{x}_j,\tau,\tau^\prime)$
%\end{equation}
The knowledge of the density matrix is useful to determine the presence of off-diagonal long-range order characterizing the superfluid phase (SF) since, in the SF phase $D_1(\vec{x}_i,\vec{x}_j)\neq 0$ for $||\vec{x}_i-\vec{x}_j||\rightarrow \infty$. The ability to detect the presence of off-diagonal long range order is  essential to study transitions from insulating to SF phases. As we shall discuss below, detecting more exotic SF phases requires the knowledge of the N-body density matrix.
% in many-body quantum systems where the transitions from localized to delocalized phases play a prominent role.

%Off-diagonal long-range order is the order featuring superfluid phases. It is the fingerprint of high delocalization of particles, and is indeed defined as $D_1(\vec{x}_i,\vec{x}_j)\neq 0$ for $||\vec{x}_i-\vec{x}_j||\leq L$, where $L$ is the system size (i.e. large probability to create and destroy particles at large distances). The ability to check the presence of off-diagonal long range order is therefore essential in the study of phase transitions in many-body quantum systems where the transitions from localized to delocalized phases play a prominent role. %\rd{something about U(1) symmetry breaking?}
%vital for the the detection of quantum-phases such as superfluids, Mott-Insulators,Supercounterflows\dots\cite{kuklovSCF,kuklovSCF2,Barbara2009,Marco1,Marco2,ourpap}.

\subsection{$N$-Body density-matrix} \label{DM}
\begin{figure}
\centering
\includegraphics[width=\columnwidth]{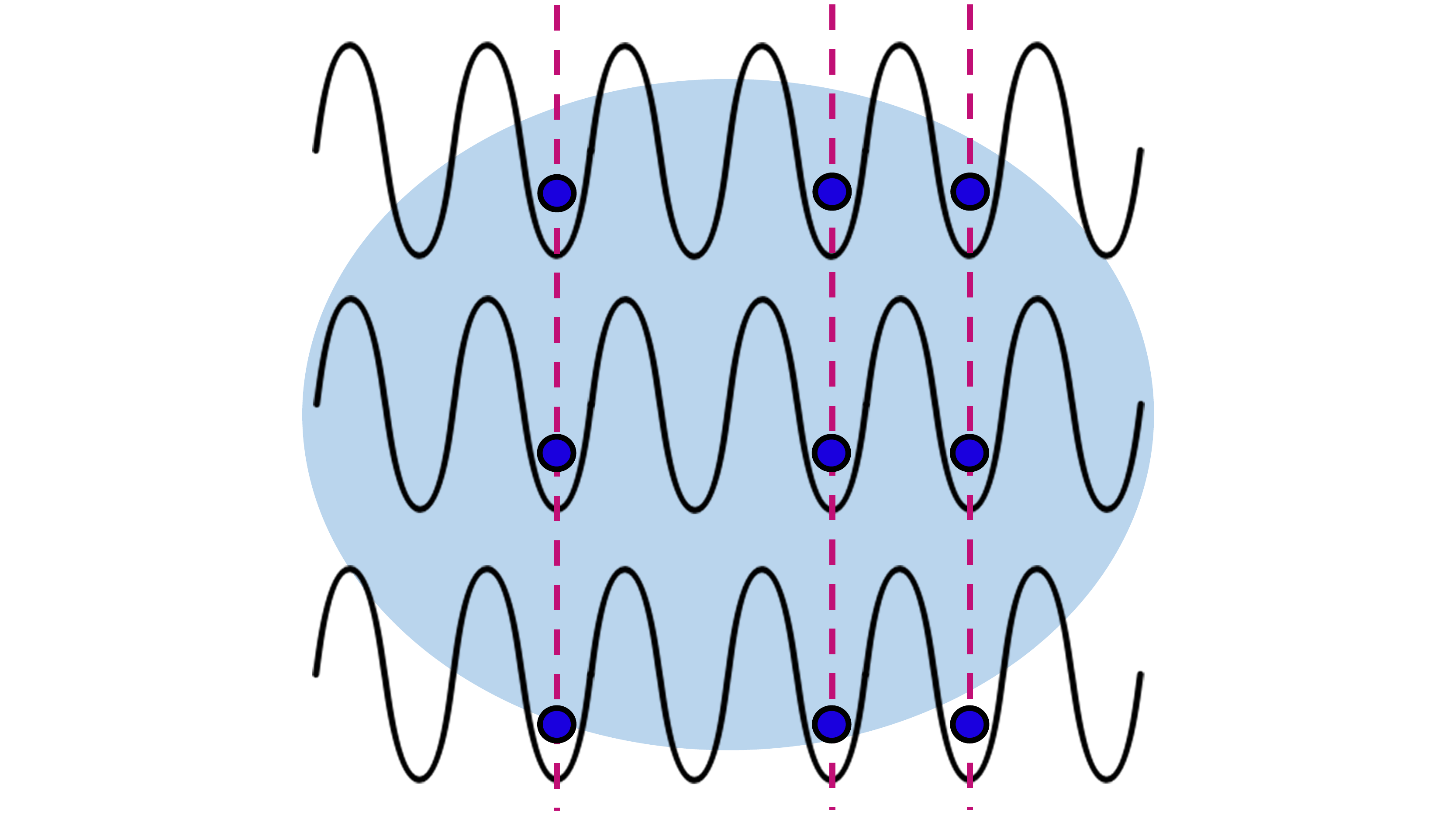}
\caption{(Color online) This figure is taken from \cite{Barbara2014}. Sketch of a SF phase of multimers. Dashed-purple lines indicate the multimers, a macroscopic bound-state among particles belonging to different layers; the extended-grey cloud signifies the delocalization of multimers over the entire lattice.}
\label{figmult}
\end{figure}

In some cases, the knowledge of the $1$-body density matrix is not sufficient to study and understand the quantum phases stabilized by the Hamiltonian. Depending on the nature of the problem and the complexity of the interaction, many-body quantum systems can exhibit quantum-phases where the correlation among many bodies plays a prominent role.
Therefore, in order to be able to fully understand the phase diagram of these systems, it is necessary to have the ability to
compute the $N$-body density matrix.
For example, systems of bosons trapped in a stack of coupled layers or systems of several interacting atomic species, can stabilize SF phases of multimers. Multimers are macroscopic multi-bound states formed by elementary particles, e.g., bound-states of particles belonging to different layers or of particles belonging to different atomic species. In FIG. \ref{figmult} we sketch the SF phase of multimers in the case of bosons trapped in a stack of $N$ optical lattice layers, where tunneling between layers is not allowed (see Section \ref{Ndist} for details). Dashed-purple lines indicate the multimers while the extended cloud underlines the delocalization of multimers over the entire lattice due to the SF phase.
SF phases of multimers are characterized by non-trivial properties of the $N$-body density matrix, where $N$ is the number of particles constituting the multimer.
%The knowledge of the $1$-body density matrix is not sufficient to grasp the complexity of certain more exotic phases \bl{such us}. For example, systems of bosons trapped in a stack of coupled layers or systems of several interacting atomic species, can stabilize SF phases of multimers. Multimers are multi-bound complexes formed by elementary constituents. Depending on the specifcs of the system, these elementary constituents may be distinguishable or indistinguishable particles or a mix of the two. SF phases of multimers are characterized by non-trivial properties of the $N$-body density matrix, where N is the number of particles constituting the multimer.
One of the goals of this paper is to show how the Worm algorithm can be generalized in order to calculate the $N$-body density matrix.

%\bl{%Depending on the application we can distinguish between two main class of lattice systems. On one hand, we may have systems that are composed by..
%Many-body quantum system may be composed, in some occasions, by distinguishable parts. For example a stack of $N$ optical-layers with no hopping among layers, or a mixture of $N$ different atomic species in the same optical lattice, are systems in which it is always possible to distinguish particles of each different layer or species. On the other hand, may also find systems where the particles are always totally indistinguishable, an example is the case of a stack of $N$ optical layers where the hopping between layers is allowed.}

Depending on the specifics of the system, the particles  in the multimer may be distinguishable or indistinguishable or a mix of the two.
Multimers made by $N$ distinguishable particles may form when the system considered is composed by $N$ distinguishable subsystems, for example a gas of N different interacting atomic species. In this case, there exists a set of $N$ distinguishable creation (annihilation) operators $a^\dag_{\alpha_k,i_k}$ ($a_{\alpha_k,i_k}$) labeled by index $\alpha_k=1...N$.  Indexes $\alpha_k$ refer to, e.g., different layers or different components. On the other hand, when multimers are made by $N$ indistinguishable particles, the index $\alpha_k$ can only assume a single value and we will therefore drop it in the notation.
%, they are all described by the same creation (annihilation) operators $a^\dag_{i_k}$ ($a_{i_k}$).

The $N$-body density operator can be defined as
\begin{equation}
D_N(\vec{x}_1,...,\vec{x}_N,\vec{x}^\prime_1,...,\vec{x}^\prime_N;\vec{\tau},\vec{\tau}\,^\prime)=
\langle A^\dag(\vec{x}_1,...,\vec{x}_N;\vec{\tau}) A(\vec{x}^\prime_1,...,\vec{x}^\prime_N;\vec{\tau}\,^\prime)\rangle,
\label{Nbody}
\end{equation}
where $\vec{\tau}=(\tau_1,...,\tau_N)$, and the operator
\begin{equation}
A(\vec{x}_1,...,\vec{x}_N;\vec{\tau}) =a_{\alpha_1,i_1}(\tau_1) ... a_{\alpha_N,i_N}(\tau_N),
\end{equation}
for the distinguishable case, and
\begin{equation}
A(\vec{x}_1,...,\vec{x}_N;\vec{\tau})=a_{i_1}(\tau_1) ... a_{i_N}(\tau_N),
\end{equation}
for the indistinguishable case.
Operators $a_{\alpha_k,i_k}(\tau_k)$ ($a^\dag_{\alpha_k,i_k}(\tau_k)$) destroy (create) a particle of type $\alpha_k$ on lattice site $i_k$ at imaginary-time $\tau_k$.
The expectation value of the $N$-body density operator in Eq. (\ref{Nbody}) represents the amplitude of the process of destroying $N$ particles in positions described by the set of coordinates $\vec{x}^\prime_1,...,\vec{x}^\prime_N$ at imaginary-times $\tau_1,...,\tau_N$, and creating them at positions described by the set of coordinates $\vec{x}_1,...,\vec{x}_N$ at imaginary-times $\tau^\prime_1,...,\tau^\prime_N$.

\begin{figure}
\centering
\includegraphics[width=0.8\columnwidth]{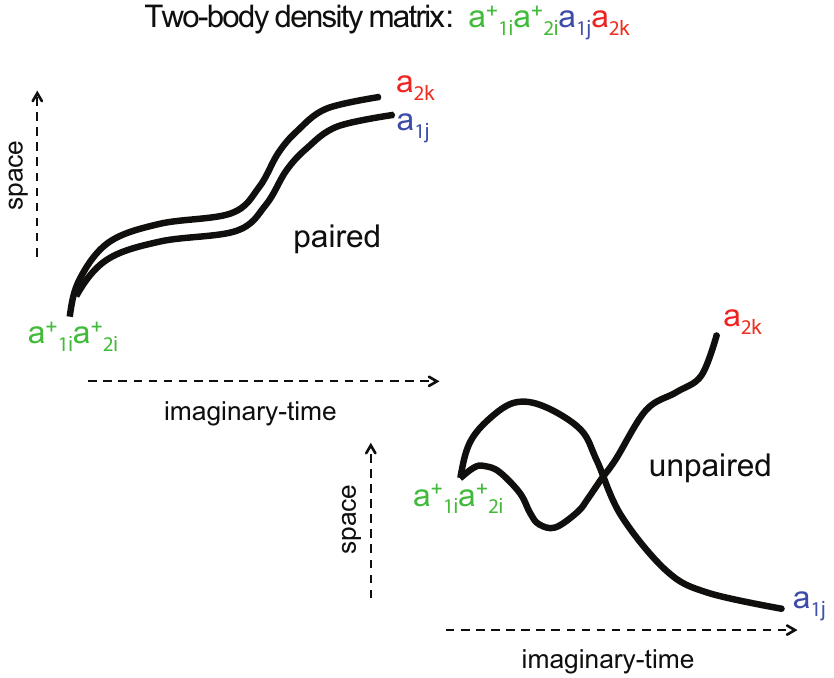}
\caption{(Color online) Sketch of the properties of the two-body density matrix for the case of $N=2$ distinguishable fields. Top:
the short-range nature of the density matrix with respect to $|\vec{x}^\prime_j - \vec{x}^\prime_k|$ is an indication that pairs are stabilized.
Bottom: density matrix in the case of independent condensates for each field (i.e. component/layer).}
\label{fig2}
\end{figure}

The features of the $N$-body density matrix give information on the quantum phase of the system. As an example, let's consider bosons trapped in a stack of $N$ layers with particle tunneling between layers suppressed and an attractive interaction between particles belonging to adjacent layers (see Section \ref{Ndist} for details). In this system, the attractive interaction is responsible for multimer formation, and because inter-layer tunneling is suppressed, multimers are made of distinguishable particles. In the ground state, a SF phase of multimers associated to a condensate of multimers (see FIG. \ref{figmult}) is stabilized.  In this phase, the corresponding density matrix $D_N$ is (i) short-ranged with respect to relative distances of the first and second set of $N$ coordinates, that is:
\begin{equation}
\iint d\vec{\tau}d\vec{\tau}\,^\prime D_N(\vec{x}_1,...,\vec{x}_N,\vec{x}^\prime_1,...,\vec{x}^\prime_N;\vec{\tau},\vec{\tau}\,^\prime) \sim e^{-\frac{|\vec{x}_m - \vec{x}_n|}{\xi}}
\end{equation}
and
\begin{equation}
\iint d\vec{\tau}d\vec{\tau}\,^\prime D_N(\vec{x}_1,...,\vec{x}_N,\vec{x}^\prime_1,...,\vec{x}^\prime_N;\vec{\tau},\vec{\tau}\,^\prime) \sim e^{-\frac{|\vec{x}^\prime_m - \vec{x}^\prime_n|}{\xi}}
\end{equation}
with $\xi \sim 1$, $\forall m,n = 1,...,N$, and (ii) long-ranged (or quasi long-ranged for the case of one-dimensional layers) with respect to
the distance between the centers of mass $|\vec{R}_{cm} - \vec{R}_{cm}^\prime|$ where
\begin{equation}
\vec{R}_{cm}= [\vec{x}_1+...+\vec{x}_N]/N,
\end{equation}
and
\begin{equation}
\vec{R}_{cm}^\prime= [\vec{x}^\prime_1+...+\vec{x}^\prime_N]/N.
\end{equation}
At the same time, all other $D_P$ ($P \neq N$) will be short-ranged
with respect to $|\vec{R}_{cm} - \vec{R}_{cm}^\prime|$. On the other hand, when particles in a given layer condense independently, the one-body density matrix $D_1(\vec{x}_\alpha;\vec{x}^\prime_\alpha)$ of each layer $\alpha$ ($\alpha = 1, ...,N$), will feature standard off-diagonal
long-range order, with all $D_M$, $M \leq N$ trivially long-ranged as they can be factorized into products of $D_1$. However, $D_M$ will no longer be short-ranged with respect to relative distances in each set of $M$ coordinates. A sketch of the two-body density matrix in the case
of $N = 2$ layers and with $\vec{x}_1=\vec{x}_2$  is shown in Fig. \ref{fig2}. The top panel corresponds to a condensate of pairs (the long- and short-range properties of $D_2$ are explicitly stated in the figure) while the bottom panel corresponds to independent condensates.

\section{Multiworm Algorithm}

%To be able to generalize the Worm-algorithm Quantum Monte Carlo to achieve the computation of the $N$-body density matrix, the Monte Carlo algorithm has to be ergodic and have access to the full statistics of $D_{N}$. This can be achieved by introducing in the updating procedure the possibility to have up to $N$ independent worms acting on the configuration space. In that case drawing and erasing updating procedures from the worms'ends will be directly linked to the critical modes, long-range order of $D_N$, hence ensuring efficiency and ergodicity.
In order to have access to the $N$-body density matrix, $N$ worms must be present in the configuration.
%As mentioned in the previous section,  the system may or not be composed of distinguishable \rd{subsystems}.
In the most general case, the $N$-body density-matrix involves the study of correlations of $N=N_{1}+...+N_{{N_D}}$ particles, where $N_D$ is the number of distinguishable particle-types in the multimer and $N_{\alpha}$, with $\alpha=1,...,N_D$, is the number of indistinguishable particles of type $\alpha$.
Configurations with $N$ worms can be generated by generalizing the source term in Eq. (\ref{source1}) as follows:
%is convenient to introduce a general method, able to generate in the same configuration both $N_I$ indistinguishable worms, and $N_D$ distinguishable worms.
%
%As seen in the previous section, to be able to generate configurations with more than one worm, it is sufficient to relax the restriction on the expansion of the off-diagonal term $\hat{H}_1(\tau) + \hat{Q}$, allowing the possibility to have multiple $a^\dag_i(\tau)$ and $a_j(\tau)$ pairs in the same configuration.
%However, doing so would allow us to generate only multiple worms of the same ``kind'' (suitable only for the study of systems with indistinguishable particles). In order to be able to study system with distinguishable particles, the possibility to generate worms of different kind must be allowed. This can be achieved by considering a generalized version of the source term (\ref{source1}):
\begin{equation}
\hat{Q}(\tau)=\sum_\alpha^{N_D} \sum_{i}[ a_{\alpha,i}^{\dag}(\tau) + a_{\alpha, i}(\tau)].
\end{equation}
%where $\alpha$ labels different distinguishable worms in the configuration.
Upon expanding $\mathbf{\hat{T}}e^{-\int_0^\beta\hat{H}_1(\tau) + \hat{Q}(\tau)\, d\tau}$ as described in \ref{app}, configurations with both distinguishable and indistinguishable worms are generated. For simplicity, we sample configurations with $N_\alpha$ worms of type $\alpha$ and neglect configurations with a number of worms $M_\alpha\ne N_\alpha$. This is enough to ensure ergodicity of the algorithm when a condensate of multimers is stabilized. To be more specific, the original worm algorithm lacks of ergodicity when used to simulate quantum phases featuring the appearance of multimers and their condensation because it is unable to generate configurations relevant to this phase, i.e., configurations where the off-diagonal many-body correlations described in Section \ref{DM} are present.%fig worms
\begin{figure}
\centering
\includegraphics[width=0.8\columnwidth]{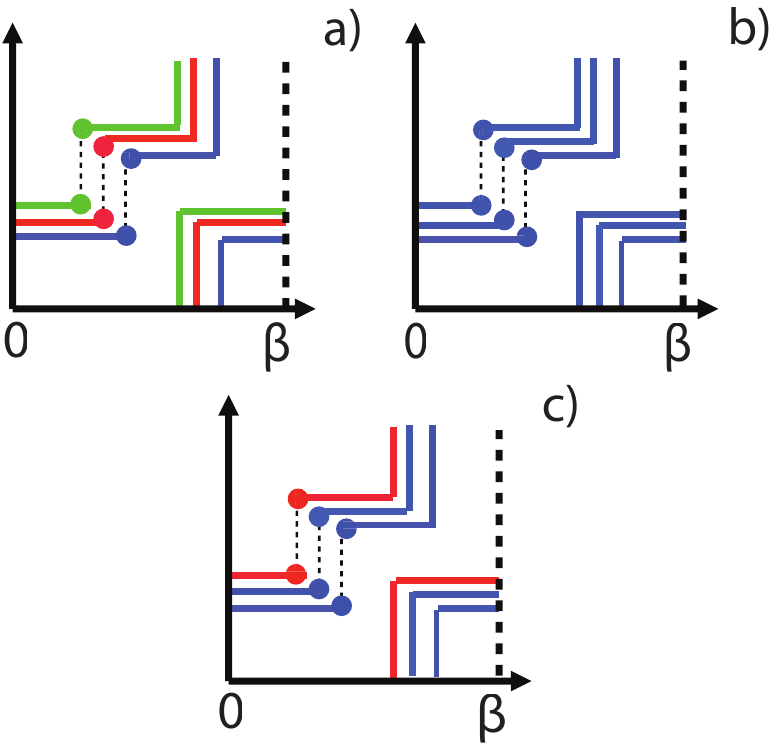}%worms.pdf
\caption{(Color online) Sketch of three possible scenarios of a multiworm configuration (we omit all other worldlines): (a) three distinguishable worms labelled with green, red and blue colors, (b) three indistinguishable worms, (c) distinguishable and indistinguishable worms.}
\label{worms}
\end{figure}
In FIG. \ref{worms} we show a sketch of three possible scenarios of a multiworm configuration (we omit all other worldlines): (a) three distinguishable worms labelled with green, red and blue colors, (b) three indistinguishable worms, (c) distinguishable and indistinguishable worms.

In order to better emphasize the difference between distinguishable and indistinguishable cases let us consider the differences in Fock states in the two cases.
If the system is composed by $N_D$-distinguishable layer or atomic species, the total Fock state $|\Psi_{N_D}\rangle$ would be the tensor product of all the $N_D$ Fock states representing the state of each distinguishable layer or component $\alpha$.
\begin{equation}
|\Psi_{N_D}\rangle= |...n_{i_1}...\rangle_1 \otimes...\otimes|...n_{i_\alpha}...\rangle_\alpha\otimes...\otimes|...n_{i_{N_D}}...\rangle_{N_D}
\end{equation}
where $|...n_{i_\alpha}...\rangle_\alpha$ is the Fock state of the layer (species) $\alpha$, and $n_{i_\alpha}$ represents the $i$-site occupation number of that layer (species).
On the other hand, when the particles are indistinguishable, the Fock state of the system is just given by $|\Psi\rangle= |...n_{i}...\rangle$.

For the sake of efficiency, creation (or annihilation) operators can be artificially kept together in space and imaginary-time by means of a weight $w\sim \exp[- \sum_{m,n}^{N} (|\vec{x}_m - \vec{x}_n|/\xi + |\tau_m - \tau_n|/\xi_\tau)]$, where $N$ is the total number of worms and $\xi$, $\xi_\tau$ are chosen in order to maximize efficiency. Clearly, expectation values have to be calculated accordingly: $\langle O\rangle= \frac{\sum_\nu O_\nu D^\nu_{N} w^\nu}{\mathcal{Z}}$  where operator $O$ describes some physical observable, $\nu$ is the generic index labeling configurations, $D^\nu_N$ is the value of the  $N$-body density matrix in configuration $\nu$, $w$ is the artificial weight, and $\mathcal{Z} = \sum_\nu D^\nu_N w^\nu$ is the normalization.

\section{Dipolar-bosons in a stack of $N$ 1D-layers}

\begin{figure}
\centering
\includegraphics[width=\columnwidth]{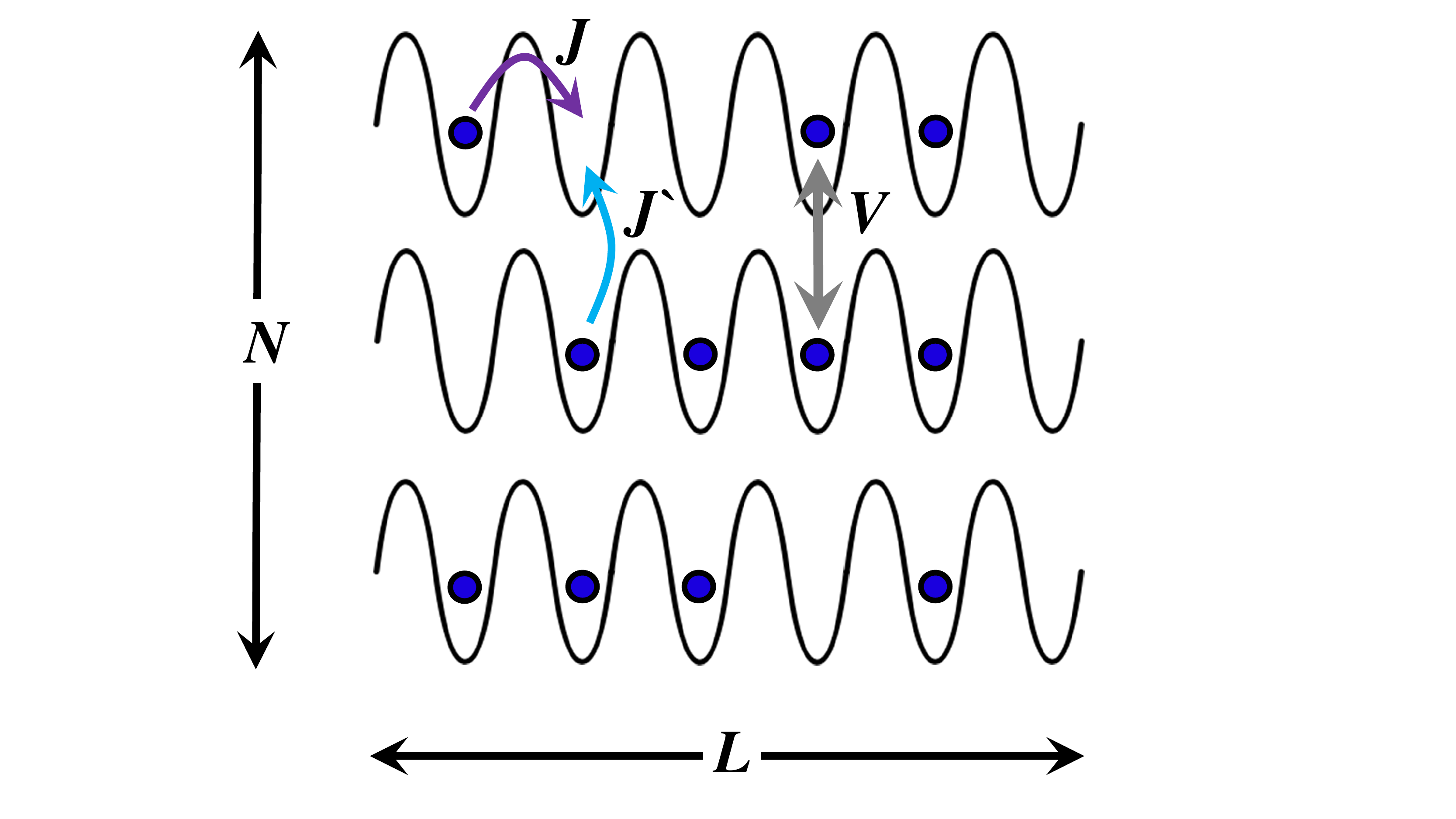}
\caption{(Color online) Sketch of the physical system consisting of N one-dimensional optical lattice layers of size $L$. $J$ is the amplitude of hopping between lattice sites on the same layer,  $J^\prime$ is the amplitude of hopping between lattice sites on different layers, and $V$ is the attractive part of the dipolar interaction between particles on different layers.}
\label{fig3}
\end{figure}

In this Section we consider two examples in which the Multiworm algorithm must be used to assure ergodicity. We study a dipolar gas of hard-core bosons trapped in a stack of $M$ one-dimensional layers. The dipole moment of each boson is aligned perpendicular to the layers and lies within the plane of the one-dimensional layers so that particles sitting on top of each other attract, while particle sitting next to each other repel. For simplicity, in the following, we cutoff the interaction so that only attraction between particles sitting on top of each other is considered.
The system is described by the Hamiltonian:
\begin{equation}
H=-J\sum_{\alpha, \langle i,j\rangle}a^\dag_{\alpha i}a_{\alpha j} - J^\prime\sum_{i, \langle\alpha ,\beta \rangle}a^\dag_{\alpha i}a_{\beta i}
- V\sum_{i, \langle\alpha ,\beta \rangle}  n_{\alpha i}n_{\beta i} - \sum_{\alpha,i} \mu_\alpha n_{i_\alpha},
 \label{Hlay}
\end{equation}
here, indexes $\alpha,\beta=1, ..., M$ label the layers, while indexes $i,j=1, ..., L$ label lattice sites within each layer; $J$ is the amplitude of hopping between lattice sites within the same layer $\alpha$, $J^\prime$ is the amplitude of hopping between different layers $\alpha$ and $\beta$; $V$ is the attractive part of the dipolar interaction, and $\mu_\alpha$ the chemical potential of the layer $\alpha$. Symbols $\langle i,j\rangle$ and $\langle\alpha ,\beta\rangle$ refers to sum over nearest neighboring sites and layers respectively. A sketch of the system with a visual explanation of the different terms in the Hamiltonian is pictured in Fig. \ref{fig3}.

%Notice that $J^\prime=0$ ($J^\prime\neq0$) corresponds to the case of $N$-distinguishable ($N$-indistinguishable) layers. %need more explanation?
Notice that if the hopping between layers is suppressed ($J^\prime=0$), it is possible to distinguish between particles on different layers. On the conrary,  if $J^\prime\neq0$, particles are all indistinguishable.
%Hence, in this system $J^\prime=0$ and $J^\prime\neq0$ correspond to the $N$-distinguishable and $N$-indistinguishable cases respectively.

In the following, we will consider two applications of the Multiworm Algorithm: a system of $M$ identical layers (\ref{Ndist}) with no hopping between layers, and a system of $M$ identical layers with finite hopping between layers (\ref{Nind}). %, and a system of $M$ layers in the presence of population imbalance (\ref{Nhybr}).

\subsection{N-indistinguishable worms}\label{Nind}

In this section we test the Multiworm algorithm on a stack of $M=N$ identical layers. For simplicity we set $J=J'$. The system is described by the Hamiltonian
\begin{equation}
H=-J\sum_{\alpha, \langle i,j\rangle}a^\dag_{\alpha i}a_{\alpha j} - J\sum_{i, \langle\alpha ,\beta \rangle}a^\dag_{\alpha i}a_{\beta i}
- V \sum_{i, <\alpha ,\beta >}n_{\alpha i}n_{\beta i} - \mu\sum_{\alpha,i} n_{\alpha\,i}
\end{equation}
Note that $\mu_\alpha=\mu$ ensures that particles density is the same on each layer. We consider periodic boundary conditions both in the direction along the layers and perpendicular to them.

We compute the ground state N-body density matrix $D_N$ and study the formation of a condensate of multimers associated to a SF phase of multimers. Multimers are formed by N indistinguishable particles where each particle belongs to a different layer. In particular, we study the transition from N independent superfluids-- one on each layer-- to a composite superfluid phase (superfluid of multimers). The latter seems to be stabilized for strong enough dipolar interaction $V/J$ though further analysis is needed to confirm it. The study is carried out for both $N=2$ and $N\geq 3$ layers, for different system sizes $L$ and different densities.

\begin{figure}
\centering
\begin{minipage}{0.48\textwidth}
\includegraphics[width=\textwidth]{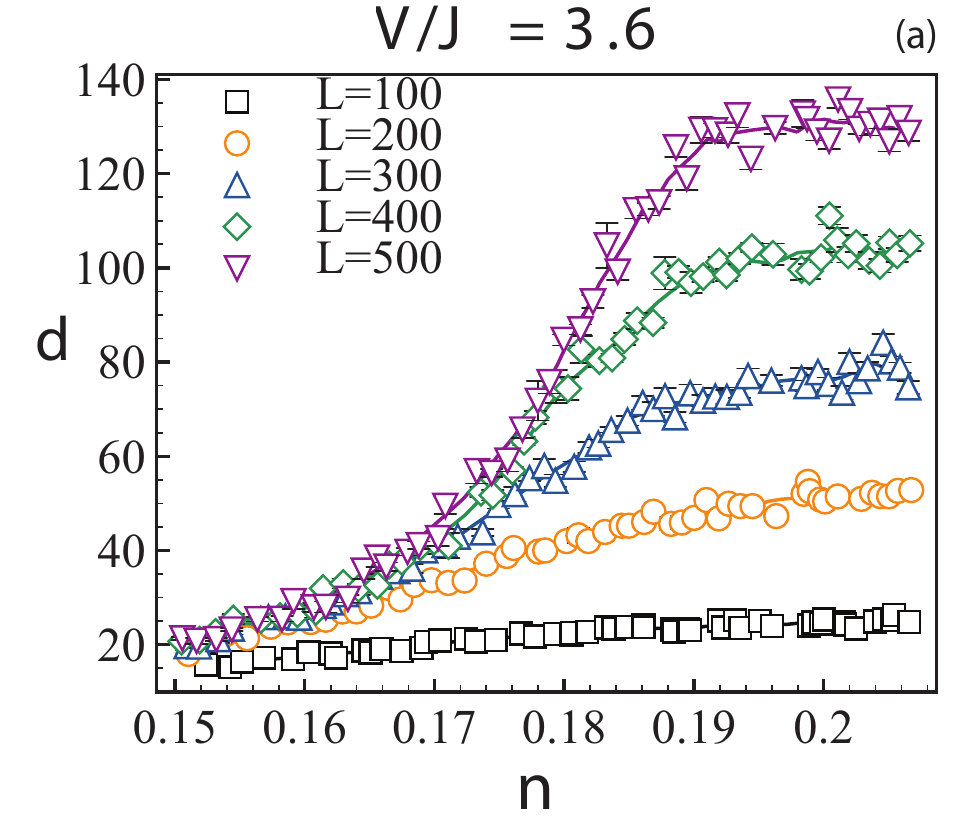}
\end{minipage}
\begin{minipage}{0.5\textwidth}
\includegraphics[width=\textwidth]{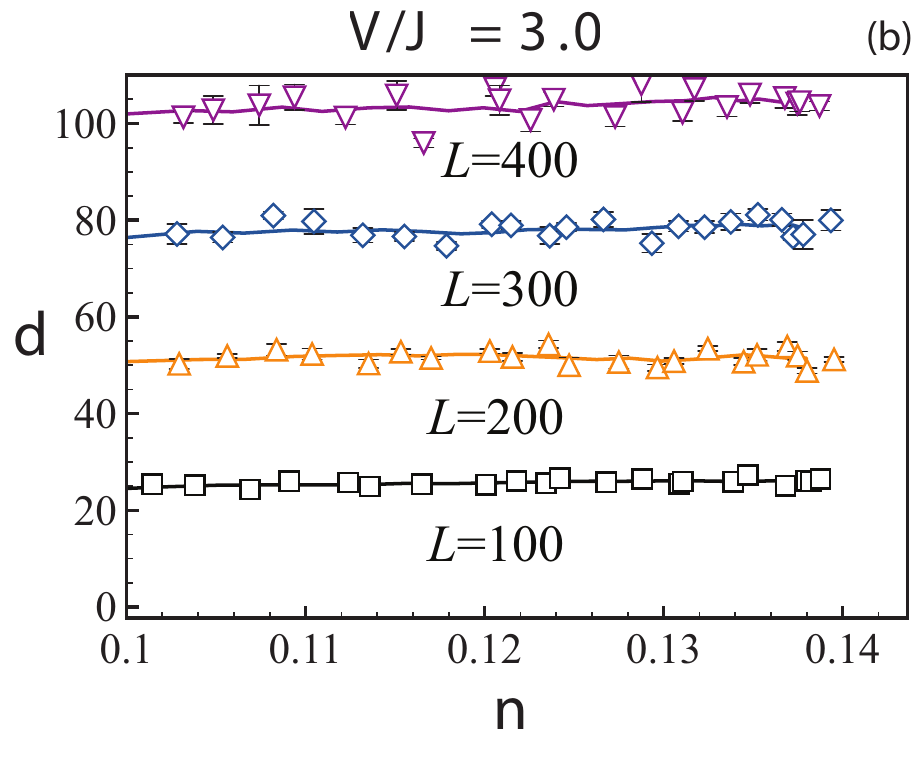}
\end{minipage}
\caption{(Color online) Average distance $d$ between worms' ends calculated according to Eq. (\ref{d_avg}) as a function of  filling $n$, for $N=2$ layers and lattice sizes $L=100, 200, 300, 400, 500$ (squares, circles, up triangles, diamonds, down triangles respectively). (a) $V/J=3.6$, a size-independent $d$ at lower densities implies that a pair-superfluid is stabilized; (b) $V/J=3.0$, a constant $d\approx \frac{L}{4}$ implies that pair-superfluidity is not stabilized at any density.}
\label{fig_akb12}
\end{figure}

We first study the case of $N=2$ layers. We refer to the superfluid phase of dimers as pair-SF (PSF). We use the following Monte Carlo observable:
\begin{equation}
d=\iint d\vec{x}_1 d\vec{x}_2 \,|\vec{x}_2 - \vec{x}_1|\, f(\vec{x}_1,\vec{x}_2)\label{d_avg}\; ,
\end{equation}
where $| \cdot |$ is the standard euclidean distance, and
\begin{equation}
f(\vec{x}_1,\vec{x}_2)=\iiiint d\vec{\tau}d\vec{\tau}^{\,\prime}
 d\vec{x}_1^{\,\prime} d\vec{x}_2^{\,\prime}\, D_2(\vec{x}_1,\vec{x}_2;\vec{x}_1^{\,\prime},\vec{x}_2^{\,\prime};\vec{\tau},\vec{\tau}^{\,\prime})\label{f12}
\end{equation}
is the probability to find the two worms' ends in positions $\vec{x}_1$ and $\vec{x}_2$ respectively. Observable in Eq. (\ref{d_avg}) represents the average distance between the worms' ends (or equivalently, between the pair of annihilation or creation operators in $D_2$).
PSF appears for strong enough $V/J$ and small enough filling factor $n=N_p/L$, where $N_p$ is the number of particles on each layer. This is shown in FIG. \ref{fig_akb12} (a) where we plot $d$ at fixed $V/J=3.6$ and system sizes $L=100, 200, 300, 400, 500$ (squares, circles, up triangles, diamonds, down triangles respectively). We notice that, for fillings $n\lesssim 0.18$,  the average distance $d$ between worms' ends drops significantly with respect to the asymptotic constant value $d\approx L/4$ and becomes system size independent. A small, size-independent $d$ reflects the short-range nature of $D_2$ with respect to relative distances in each set of coordinates $\{x_i\}$ and $\{x'_i\}$ as discussed in Section~\ref{DM}.  PSF is destabilized as filling factor is increased. This is a purely many-body effect. As $n$ increases, particle-exchanges between dimers are favored and a transition to two independent SF (2SF) --one on each layer-- seems to occur. The lower the interaction strength, the lower the density at which PSF may be observed. In FIG. \ref{fig_akb12} (b), we plot $d$ for fixed dipolar interaction $V/J=3.0$ and system sizes $L=100, 200, 300, 400$ (squares, up triangles, diamonds, down triangles respectively). We did not find any evidence of PSF phase for density as low as $n\sim 0.1$ as an approximately constant $d\approx L/4$ upon varying density demonstrates. This corresponds to a 2SF phase.
%the case of unpaired superfluids, in which the two-worms explore independently \rd{(and uniformly??)} the configuration space.
This phase features long-range of $D_2$ with respect to both $|\vec{x}_i - \vec{x}_j |$ and $|\vec{x}_i - \vec{x}_i^{\,\prime} |$.
%On the other hand, when the average distance is small, the two worms tends to explore the configuration space coupled between each other, evidencing the presence of a bound-state between particles, i.e. paired superfluidity.
For the $2$ layer case we observe a critical $V_c/J\approx 3.1$ between the two phases. Overall, for the $2$ layer case, by simulating system sizes up to L=500, we have observed a PSF phase at low enough density and interaction values $V/J\gtrsim 3.1$.

\begin{figure}
\centering
\includegraphics[width=0.8\textwidth]{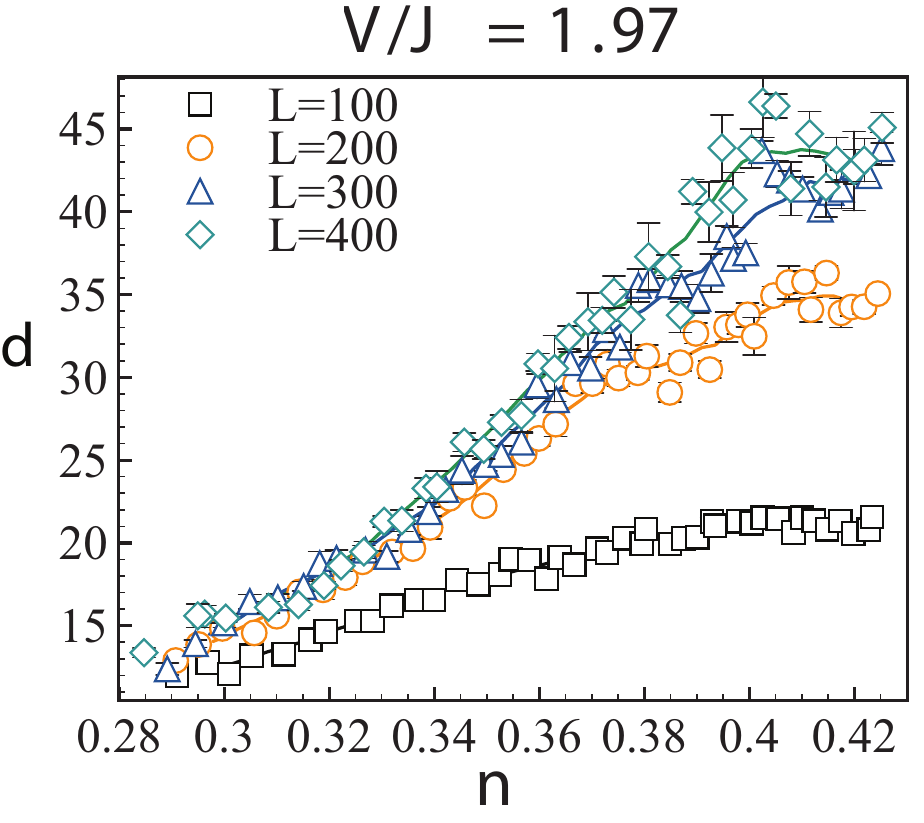}
\caption{(Color online) Average distance $d$ between worms' ends calculated according to Eq. (\ref{f12_3}) as a function of  filling $n$, for $N=3$ layers, $V/J=1.97$ and lattice sizes $L=100, 200, 300, 400$ (squares, circles, triangles and diamonds respectively). Multimers are formed up to density $n\sim 0.35$ as a size-independent $d$ implies.}
\label{fig_akb3}
\end{figure}

Similar results are also found for a number of layers $N>2$. Our main finding is that, for $N>2$, multimers are stabilized at lower interaction strength and survives for larger densities. For example, for $N=3$ and $V/J= 1.97$, we find that, for the system sized considered here, multimers are formed up to density as large as $n\sim 0.35$. This is shown in FIG. \ref{fig_akb3} where we plot $d$ as a function of $n$ for system sizes $L=100, 200, 300, 400$ (squares, circles, triangles and diamonds respectively). Here $d$ is calculated according to Eq. (\ref{d_avg}), generalizing the definition of the probability distribution in Eq. (\ref{f12}) as
\begin{equation}
f(\vec{x}_1,\vec{x}_2)=\iiiint d\vec{\tau}d\vec{\tau}^{\,\prime}
 d\vec{x}_3 d\vec{x}_1^{\,\prime} d\vec{x}_2^{\,\prime} d\vec{x}_3^{\,\prime}\, D_3(\vec{x}_1,\vec{x}_2,\vec{x}_3;\vec{x}_1^{\,\prime},\vec{x}_2^{\,\prime},\vec{x}_3^{\,\prime};\vec{\tau},\vec{\tau}^{\,\prime})\label{f12_3}
\end{equation}
where invariance under the exchange of $\vec{x}_1$, $\vec{x}_2$, and $\vec{x}_3$ has been verified. Finally, we find very similar results for a number of layers $N>3$. This may be  due to the fact we are considering only nearest-neighbor interactions.

% \subsection{Hybrid case}\label{Nhybr}
% \dots to be written soon\dots waiting for the results\dots

\subsection{N-distinguishable worms}\label{Ndist}
In this section we summarize some of the results from B. Capogrosso-Sansone \emph{et. al.} in \cite{Barbara2014} in which a Multiworm algorithm has been used to study a stack of $M=N$ one-dimensional layers where the hopping among layers has been suppressed ($J^\prime=0$). The system is described by the Hamiltonian
\begin{equation}
H=-J\sum_{\alpha, \langle i,j\rangle}a^\dag_{\alpha i}a_{\alpha j}
- V\sum_{i, \langle\alpha ,\beta \rangle}  n_{\alpha i}n_{\beta i} - \sum_{\alpha,i}  \mu_\alpha n_{i_\alpha},
 \label{HNdist}
\end{equation}
where $J$ is the hopping amplitude, and $V$ is the attractive dipolar interaction among nearest layer. Note that $\mu_\alpha=\mu$ ensures that particles density is the same on each layer $n=N_p/L$, where $N_p$ is the number of particles on each layer. We consider periodic boundary conditions along the layers and in the direction perpendicular to layers.

\begin{figure}
\centering
\includegraphics[width=0.8\columnwidth]{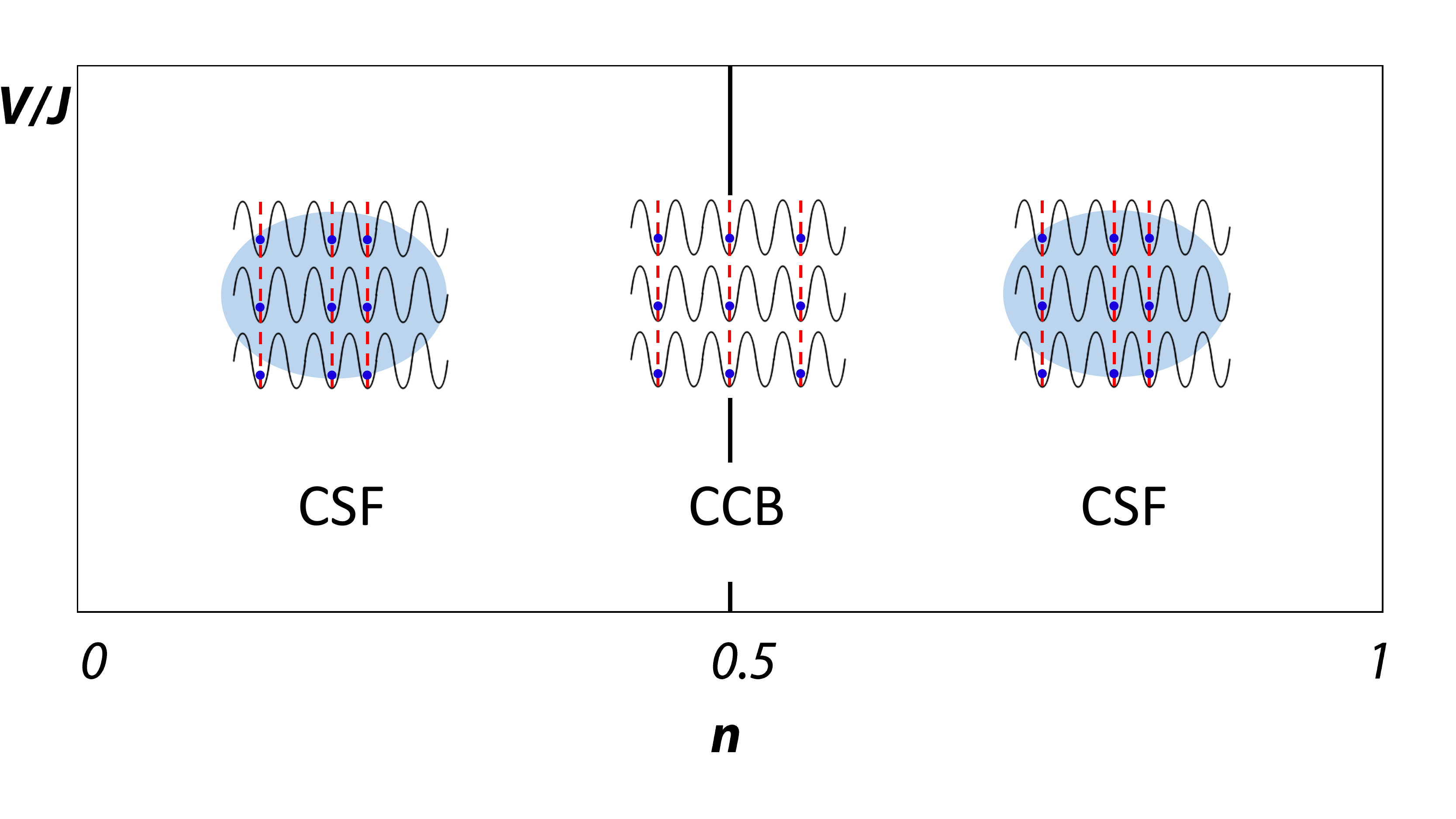}
\caption{ Sketch of the quantum phases stabilized by model~\ref{HNdist} for $N>2$ one-dimensional layers. $V$ is the attractive dipolar interaction, $J$ is the hopping amplitudes, and $n$ is the filling on each layer. The system is in a chain-checkerboard phase at $n=0.5$, and in a chain-superfluid phase elsewhere.}
\label{phDiagNdist}
\end{figure}

%The symbol $\langle\cdot,\cdot\rangle$ refers to summation over nearest neighbors sites and layers.
%The layer's chemical potential has been set to $\mu_\alpha=\mu$ in order to achieve uniform filling $n=N_p/(L\cdot N)$ (i.e. $N_p$ number of particles).
Since hopping between layers is suppressed, particles belonging to different layers are distinguishable. As shown in \cite{Barbara2014}, and sketched in FIG. \ref{phDiagNdist}, for $N>2$ and $V/J\ne 0$ this system undergoes a phase transition from a chain-superfluid phase (CSF) phase at generic filling, to a chain-checkerboard phase (CCB) at filling $n=0.5$. As discussed previously, a chain-superfluid is a superfluid of multimers (see FIG. \ref{figmult}), while the checkerboard phase is an insulating phase in which the multimers (``vertical chains'' of particles spanning across the layers) arrange themselves in a checkerboard fashion, i. e. each occupied site is surrounded by unoccupied neighbors (see FIG. \ref{phDiagNdist}).

\begin{figure}
\centering
\includegraphics[width=0.8\columnwidth]{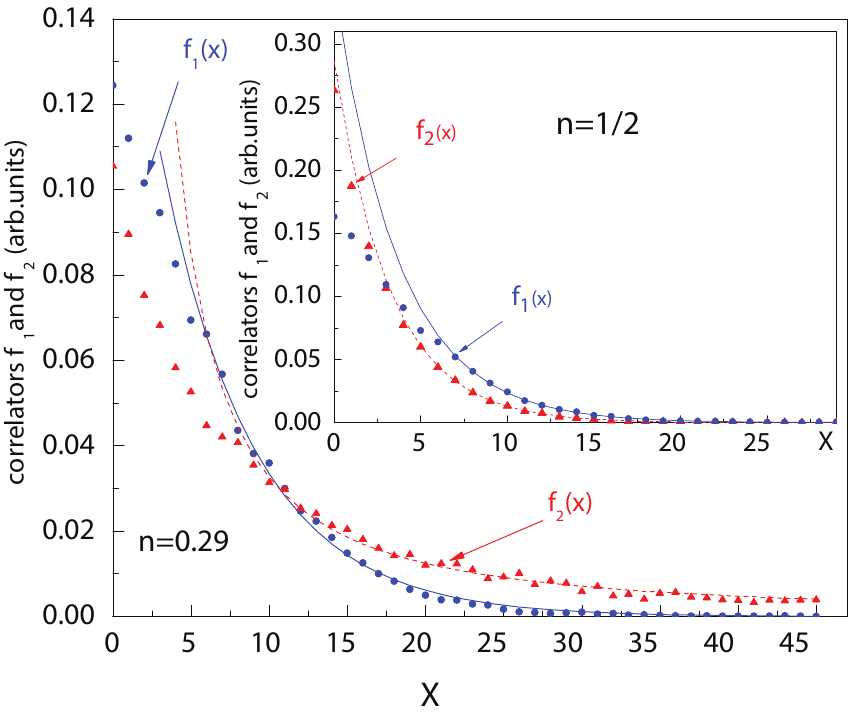}
\caption{(Color online) This figure is taken from \cite{Barbara2014}. $f_1(X)$ and $f_2(X)$ for $n=0.29$ (main panel) and for $n=0.5$ (inset). $X=x_1^\prime-x_2^\prime$ for $f_1$, and $X=x_1-x_1^\prime$ for $f_2$. Main panel: the system is in a chain-superfluid phase since $f_1$ decays exponentially as $\sim e^{-0.169|X|}$ and $f_2$ decays algebraically as $\propto |X|^{-1.39}$. Inset: the system is in a chain-checkerboard phase as both $f_1$ and $f_2$ decay exponentially -- $f_1(X)\sim e^{-0.269|X|}$ and $f_2(X)\sim e^{-0.310|X|}$.}
\label{fNdist}
\end{figure}

Here, we show how, by studying the features of the many-body correlator $D_N$, one can infer the quantum phases stabilized by model in Eq. (\ref{HNdist}). We report results for the case of $N=3$ layers. Let us define the two quantities
\begin{equation}
f_1(x_1^\prime-x_2^\prime)\propto \int d\vec{\tau} d\vec{\tau}^\prime dx_1 dx_2 dx_3 dx_3^\prime D_3,
\end{equation}
and
\begin{equation}
f_2(x_1-x_1^\prime)\propto \int d\vec{\tau} d\vec{\tau}^\prime dx_2 dx_3 dx_2^\prime dx_3^\prime D_3,
\end{equation}
where $D_3\equiv D_3(x_1,x_2,x_3;x_1^\prime,x_2^\prime,x_3^\prime;\vec{\tau},\vec{\tau}^\prime)$ is the $3$-body density matrix between particles belonging to the three different layers. According to their definition, $f_1$ should manifest exponential decay in both CSF and CCB phases, while $f_2$ should decay exponentially in the CCB phase and algebraically in the CSF phase.
Some of our results are reported in FIG. \ref{fNdist} where we show that, for filling $n=0.29$ (main panel), the system is in a CSF since $f_1$ manifests an exponential decay as $f_1(X)\sim e^{-0.169|X|}$ while $f_2$ decays algebraically as $f_2(X)\propto |X|^{-1.39}$. On the other hand, at filling $n=0.5$ (inset), the system is in the CCB phase as both $f_1$ and $f_2$ decay exponentially: $f_1(X)\sim e^{-0.269|X|}$, and $f_2(X)\sim e^{-0.310|X|}$. These results demonstrate the effectiveness of the Multiworm algorithm to study SF phases of multimers.

\section{Conclusions}
We reviewed the theoretical formulation of path-integral Quantum Monte Carlo techniques and presented its implementation with a Multiworm algorithm suitable to study multi-component systems.
We showed how the configuration space in which the Multiworm algorithm works naturally allows for the computation of the $N$-body density matrix and many-body correlations.
%the $N$-body density matrix capable of describing many-body correlations
We applied the algorithm to dipolar lattice bosons trapped in a stack of $N$ one-dimensional layers with zero and finite particle-tunneling between adjacent layers.
We studied the N-body correlation properties of the system from which we were able to infer the quantum phases stabilized. We found that a superfluid of multimers made of inidstinguishable particles is present when particle-tunneling between layers is finite. We observed this phase at large enough dipolar interaction and at low enough densities. Similarly, when inter-layer particle-tunneling is turned off, a superfluid phase of multimers made of distinguishable particles is stabilized for any interaction strength and densities other than 0.5 where for $N>2$ a checkerboard solid is present.

In conclusions, the algorithms presented are suitable to study complex dipolar lattice bosons and multi-component bosonic systems.

{\em Acknowledgements}
This work  was  supported  by  the  NSF  (PIF-1552978).   The
computing  for  this  project  was  performed  at  the  OU
Supercomputing  Center  for  Education  and  Research
(OSCER) at the University of Oklahoma (OU).

\appendix

\section{Path-integral in continuous imaginary-time}\label{app}

Within the interaction-picture formalism the Hamiltonian is split into two parts:
\begin{equation}
\hat{H}=\hat{H}_0 + \hat{H}_1.
\end{equation}
Here $\hat{H}_0$ is the diagonal part and $\hat{H}_1$ is the off-diagonal part of the Hamiltonian in some convenient basis.
For Bose-Hubbard models (see Eq. (\ref{BH})), the diagonal part in the Fock representation is given by  $\hat{H}_0=\sum_{i,j}V_{ij}n_i n_j + \sum_i U_i n_i(n_i - 1) -\sum_i \mu_i n_i$, and the off-diagonal part is given by $\hat{H}_1=-\sum_{i,j} J_{ij} a^\dag_i a_j$.
The imaginary-time evolution operator can be expressed \cite{prkof_2} as:
\begin{equation}
e^{-\beta\hat{H}}=e^{-\beta\hat{H}_0}\cdot\mathbf{\hat{T}}e^{-\int_0^\beta\hat{H}_1(\tau)d\tau}
\end{equation}
where $\mathbf{\hat{T}}$ is the time-ordering operator, and
\begin{equation}
\hat{H}_1(\tau)= e^{\hat{H_0}\tau}\hat{H}_1e^{-\hat{H_0}\tau},\label{E1}
\end{equation}
where $\hat{H}_1=\hat{H}_1(0)$.
%, while $\hat{H}_0$ and $\hat{V}$ are the diagonal and off-diagonal part of Hamiltonian $\hat{H}$, respectively.
Within this representation, the Matsubara time evolution operator $\hat{\sigma}= \mathbf{\hat{T}}e^{-\int_0^\beta\hat{H}_1(\tau)d\tau}$ can be expanded as
\begin{equation}
\hat{\sigma}= \mathbf{\hat{T}}e^{-\int_0^\beta\hat{H}_1(\tau) d\tau}=\mathbb{1}  + \hat{\sigma}^{(1)} + \dots + \hat{\sigma}^{(n)}\label{evolexp}
%=\mathbb{1} - \int_0^\beta\hat{V}_{(\tau)}d\tau + \int_0^\beta d\tau \int_0^{\tau_2} d\tau_1 \hat{V}_{(\tau)}\hat{V}_{(\tau_1)} +\\
%+ \dots+\\
%(-1)^n \int_0^\beta d\tau_n \int_0^{\tau_{n}} d\tau_{n-1} \cdots\int_0^{\tau_{2}} d\tau_1 \hat{V}_{(\tau_{n})}\hat{V}_{(\tau_{n-1})}\cdots\hat{V}_{(\tau)}=\\
\end{equation}
where the generic, $n$-th order term, has the form
\begin{equation}
\hat{\sigma}^{(n)}
 =(-1)^n \int_0^\beta d\tau_n \cdots\int_0^{\tau_{2}} d\tau_1 \hat{H}_1(\tau_{n})\hat{H}_1(\tau_{n-1})\cdots\hat{H}_1(\tau_1).\label{nterm}
\end{equation}
Eq. (\ref{evolexp}) and Eq. (\ref{nterm}) are obtained by reformulating the imaginary-time Schrodinger equation $-\partial_\beta |\Psi(\beta)\rangle=\hat{H}_1(\beta)|\Psi(\beta)\rangle$ in the interaction picture
\begin{equation}
-\partial_\beta |\Phi(\beta)\rangle=\hat{H}_1(\beta)|\Phi(\beta)\rangle,\;\;\; \hat{H}_1(\beta)=e^{\beta\hat{H}_0}\hat{H}_1 e^{-\beta\hat{H}_0},
\end{equation}
with $|\Psi(\beta)\rangle=e^{-\beta\hat{H}_0}|\Phi(\beta)\rangle$. This, in turn, can be written in the significant integral form
\begin{equation}
|\Phi(\beta)\rangle=|\Phi(0)\rangle - \int_0^{\beta}\hat{H}_1(\tau_1)|\Phi(\tau_1)\rangle.
\label{PHI}\end{equation}
Using Eq.  (\ref{PHI}),  $|\Phi(\tau_1)\rangle$ can in turn be expressed with the same integral form. Then,
repeating this process iteratively one obtains (up to a residual contribution tending to zero for $N\rightarrow\infty$) the expression
\begin{multline}
\!\!\!\!\!|\Phi(\beta)\rangle=\Big[ \mathbb{1} + \sum_{k=1}^N(-)^k \int_0^\beta \!\!\!d\tau_n \hat{H}_1(\tau_n)\int_0^{\tau_{n}} \!\!\!d\tau_{n-1} \hat{H}_1(\tau_{n-1})\\
\dots\int_0^{\tau_{3}} \!\!\!d\tau_{2} \hat{H}_1(\tau_{2})\int_0^{\tau_{2}} \!\!\!d\tau_{1} \hat{H}_1(\tau_{1})\Big]|\Phi(0)\rangle
\end{multline}
readily providing the time-evolution operator in Eq. (\ref{evolexp}) and definition in Eq. (\ref{nterm}).

The chain of operators $\hat{H}_1(\tau_{n})\hat{H}_1(\tau_{n-1})\cdots\hat{H}_1(\tau)$ describes the evolution of the system between the imaginary time $\tau=0$ and $\tau=\beta$.
Within this formalism the trace in expression (\ref{expO}) can be rewritten as:
\begin{equation}
Tr(e^{-\beta\hat{H}}\hat{O})=\sum_{\alpha, n} \langle \alpha| e^{-\beta\hat{H}_0}\hat{\sigma}^{(n)} \hat{O} |\alpha \rangle \label{tr1}
\end{equation}
Here, $\{|\alpha\rangle\}$ and $\{E_\alpha\}$ are the eigenstates and eigenvalues of $\hat{H}_0$. By explicitly writing $\hat{\sigma}^{(n)}$ in Eq. (\ref{tr1}), it is possible to rewrite the trace into its final form shown in Eq. (\ref{pathint}).
\begin{multline}
Tr(e^{-\beta\hat{H}}\hat{O})=\sum_{\alpha, n}  \int_0^\beta d\tau_n \cdots\int_0^{\tau_{2}} d\tau_1 (-1)^n e^{-\beta E_\alpha} \times\\
\times\langle \alpha|  \hat{H}_1(\tau_{n})\hat{H}_1(\tau_{n-1})\cdots\hat{H}_1(\tau_1) |\Theta_\alpha \rangle \label{tr2}
\end{multline}
where $|\Theta_\alpha \rangle=\hat{O} |\alpha \rangle$ is the Fock state resulting from the action of operator $\hat{O}$ on the state $|\alpha\rangle$.

In the following, we further specify Eq. (\ref{tr2}) using Hamiltonian in Eq. (\ref{BH}) for the computation of the partition function in Eq. (\ref{partf}).
Notice that the partition function is just a simple case of Eq. (\ref{tr2}) in which $\hat{O}=\mathbb{1}$, the same reasoning applies straightforwardly for the computation of the expectation value of the generic observable $O$.
By inserting completeness relations $\sum_{\alpha_i} |\alpha_i\rangle\langle\alpha_i|=\mathbb{1}$ between every two consecutive ``hopping'' operators $\hat{H}_1(\tau_i)$ in Eq. (\ref{tr2}), it is possible to explicitly write the amplitude $\langle\alpha|\hat{H}_1(\tau_{n})\cdots\hat{H}_1(\tau_1) |\alpha \rangle$ as a sum of amplitudes of all the possible paths from $|\alpha\rangle$ to $|\alpha\rangle$.
%Here, each single-path is defined by $n$ intermediate steps, consisting of $n$ intermediate propagations among the different states $|\alpha_i\rangle$s, with $i=1...n$.
\begin{multline}
\langle \alpha|  \hat{H}_1(\tau_{n})\cdots\hat{H}_1(\tau_1) |\alpha \rangle=\sum_{\alpha_1,...,\alpha_{n-1}} \langle\alpha|\hat{H}_1(\tau_{n})|\alpha_{n-1}\rangle\times\\
\times\cdots\times\langle\alpha_{i}|\hat{H}_1(\tau_{i})|\alpha_{i-1}\rangle\times\cdots\times \langle\alpha_1|\hat{H}_1(\tau)|\alpha\rangle \label{pathampl}
\end{multline}
%The total amplitude of the propagation is then given by the sum of the amplitude of all the possible paths ().
The amplitude of a single path is given by the product of all the intermediate transitional amplitudes describing the propagation from the intermediate state $|\alpha_{i-1}\rangle$ at $\tau=\tau_{i-1}$, to the state $|\alpha_i\rangle$ at $\tau=\tau_{i}$.
Namely,
\begin{equation}
\langle \alpha|  \hat{H}_1(\tau_{n})\cdots\hat{H}_1(\tau_1) |\alpha \rangle=\\
\sum_{\alpha_1,...,\alpha_{n-1}} H_1^{\alpha\alpha_{n-1}}(\tau_{n})\cdots H_1^{\alpha_2\alpha_1}(\tau_{2})H_1^{\alpha_1\alpha}(\tau_1) \label{pathampl}
\end{equation}
where
\begin{equation}
H_1^{\alpha\beta}(\tau)=e^{(E_\alpha-E_\beta)\tau}\langle\alpha| H_1 |\beta\rangle,
\end{equation}
and expression (\ref{E1}) has been used.
Exploiting the properties of the hopping operator, and orthogonality among different Fock states, one can rewrite the matrix element as
\begin{equation}
\langle\alpha| H_1 |\beta\rangle=-\sum_{i,j}J_{ij}\langle\alpha| a^\dag_i a_j |\beta\rangle=-J_{ij} \sqrt{(n_i^{(\alpha)} + 1)n_j^{(\beta)}} \label{hop}
\end{equation}
where $H_1^{\alpha\beta}(\tau)\neq0$, if and only if states $|\alpha\rangle$ and $|\beta\rangle$ differ only in their occupation numbers at sites $i$, $j$ such that $n_j^{(\alpha)}=n_j^{(\beta)}-1$ and $n_i^{(\alpha)}=n_i^{(\beta)} + 1$.

The partition function is then expressed as
\begin{equation}
\mathcal{Z}=\sum_{\alpha,n}  \sum_{\{\alpha_p\}} A^n \int_0^\beta d\tau_n \cdots\int_0^{\tau_{2}} d\tau_1  \prod_{p=1}^n e^{-\beta E_{\alpha_p}(\tau_{p}-\tau_{p-1})} \label{pf2}
\end{equation}
where $\{\alpha_p\}=\alpha_1,\alpha_2,\dots\alpha_{n-1}$, and $A^n$ contains the product of square roots and hopping amplitudes of Eq. (\ref{hop}). Eq. (\ref{pf2}) can be formally rewritten as
\begin{equation}
\mathcal{Z}=\sum_{\{\nu\}} W_\nu
\end{equation}
where $\{\nu\}$ is a collection of discrete and continuous indexes, and
$W_\nu=A^n  \prod_{p=1}^n e^{-\beta E_{\alpha_p}(\tau_{p}-\tau_{p-1})}$ is the weight of each configuration.

To summarize, the expectation value of the observable $O$ and the partition function $\mathcal{Z}$ can be computed as a sum of all possible evolutions in imaginary-time from all the possible initial states $ |\Theta_\alpha \rangle$ at $\tau=0$, to the corresponding definite final state $|\alpha \rangle $ at $\tau=\beta$. These paths in imaginary-time are called \emph{configurations}. Every configuration is therefore defined by the product of ``hopping amplitudes'' $\langle \alpha| \hat{H}_{1}(\tau_{n})|\alpha_n\rangle\cdots\langle\alpha_1|\hat{H}_{1}(\tau_1)|\Theta_\alpha\rangle$ that, by definition, fixes the path in imaginary-time from state $ |\Theta_\alpha \rangle$ to the state $ |\alpha \rangle$. Notice that, for the sake of simplicity of notation, in Eq. (\ref{pathint}) we summarized the multiple sum and integrals of Eq. (\ref{tr2}-\ref{pathampl}) as $\sum_{paths} p_\alpha$, with  $p_\alpha=(-1)^n e^{-\beta E_\alpha}$.

\end{document}